# Service mining for Internet of Things

Bing Huang[?]

Springer-Verlag, Computer Science Editorial,
Tiergartenstr. 17, 69121 Heidelberg, Germany

**Abstract.** The abstract should summarize the contents of the paper and should contain at least 70 and at most 150 words. It should be written using the *abstract* environment.

**Keywords:** We would like to encourage you to list your keywords within the abstract section

## 1  Introduction

The current Internet is evolving from interconnecting computers to interconnecting things [14].There are 9 billion interconnected things and it is expected to reach 24 billion things connected by 2020.According to the GSMA, this amounts to 1.3 trillion revenue opportunities for health, automotive, utilities and consumer electronics [14].The Internet of Things (IoT)can be considered as a global network infrastructure composed of numerous connected things that rely on sensory, communication, networking, and information processing technologies [12]. Things can be a place, a physical or computational object[12].The emerging Internet of Things (IoT) bridges the gap between the physical and the digital worlds,which enables a deeper understanding of people¡s behaviors. Things are closely correlated with human behaviors, because human behaviors can be inferred by analyzing their interactions with things [5].Hence, revealing the implicit relationships among things is of paramount important in understanding human behaviors. By doing so, human behaviors can be learned and proper actions can be taken to react to their needs and even anticipate potential changes.

   Revealing things relationships is challenging for two reasons. First, things are heterogeneous in terms of functionalities and properties [3]. Second, there are little standard descriptions of things [21]. Third, things relationships are sometimes implicit and difficult to be discovered. Service paradigm may be a promising solution to support standardization of things descriptions and access methods. The service paradigm is a powerful abstraction hiding complex internal logical features of things and focusing on how things are to be accessed. In this regard, things can be abstracted as services [31].

---

? Please note that the LNCS Editorial assumes that all authors have used the western naming convention, with given names preceding surnames. This determines the structure of the names in the running heads and the author index.

The problem of discovering things relationships can be therefore formulated as discovering service relationships (i.e., service mining). We define service mining as the approach of proactively discovering interesting relationships among services. The ultimate product of service mining is a set of service composition leads. The service composition lead is essentially a group of services that are related to each other. Service mining takes a bottom-up approach. The bottom-up approach aims at discovering service composition leads without requiring a user to specify search criteria. By contrast, service composition traditionally takes a top-down approach [10].The top-down approach requires a user to provide a goal containing specific search criteria defining the exact service functionality the user expects [10]. Thus, unlike the service composition which is strictly driven by the search criteria, service mining has the potential of discovering service relationships that are unexpected by many people.

Discovering service relationships is challenging for mainly three reasons. First, things are sometimes dynamic, because they may represent different states in different contexts. The context is any information that can be used to characterize the situation of things [8]. The context includes time and location of things, things relationship to people, and things relationship to the environments [8]. Services should not only model things functionalities and properties, but also model things contexts. Current service ontologies describe services from a syntactic and semantic point of view [15][19][20]and provide little support for describing context. Second, much like molecules in the natural world where they can recognize each other and form bonds in between [10], services can also recognize each other through syntax, semantics and context. A mechanism is needed to enable automatic service recognition and to form bonds between services. Third, once the relationship between two services is identified, the relationship may expand due to the fact that the two services may relate to other services. Since the number of services may be huge, service relationships may expand uncontrollably, leading to the problem of combinatorial explosion. Fourth, discovered service relationships might be already known or useless. Evaluation methodologies are needed to differentiate interesting service relationships from those that are already known or useless.

In this paper, we propose a service mining framework that attempts to addresses the four challenges. The rest of the paper is organized as follows: Section 2 gives an overview of related work. Section 3 propose a service mining ontology model. Section 4 elaborates service mining framework. Section 5 proposes an evaluation methodology. Section 6 evaluates the proposed evaluation methodology and shows the experiment results. Section 7 concludes the paper and highlights some future work.

## Motivation

As the prevalence of IoT, an increasing number of things will be connected, creating tremendous opportunities of discovering things relationships. The opportunities of things relating to each other will be beyond people¡s imagination.

While we may not sometimes have specific search criteria in finding the implicit things relationships, being able to proactively discover these relationships is of great value. Service mining provides a means to reveal things relationships. Service mining has the potential to change the way of living with many applications in public transportation management, healthcare industry, and smart home etc. For transportation authorities, service mining can be used to find relationships between public transport infrastructures. Thus the transportation management will be improved. People could consume better public transportation services in advance. For most household occupants, discovering the implicit service relationships in homes is beneficial for saving energy and enhancing accommodation comfort. For healthcare givers, service mining enables discovering people¡s repetitive behaviors unobtrusively, thus providing better healthcare services. From a technical point of view, discovering service relationships is critical for service recommendation and service composition. Therefore, it is essential to be able to proactively discovering the implicit services relationships even when the search criteria are unknown.

## Scenario

A smart home can be defined as a dwelling in which things are connected and can cooperatively respond to people¡s needs automatically, aiming to promote living comfort, convenience, security, and entertainment [1][2]. We assume all things in smart home such as TV, DVD, fridge, door and chair (equipped with sensors) etc. are abstracted as services. Services are registered in a service repository. For simplicity, we use A.m() to refer to service operation. In our scenario, we adopt the framework proposed in [3] that abstracts things as services. In [3], there is an embedded device middleware layer locating between a physical things layer (i.e., device layer) and a service layer, as shown at the bottom of Fig.1. The embedded device middleware abstracts the feature of things and exposes things features as control APIs. The service layer aggregates the control APIs according to the logical features of things. In this regard, things in smart home are abstracted as services and can be controlled or observed through invoking corresponding services. For example, the TV can be controlled through the TV service which contains a set of operations such as TV.ON(), TV.setVolume(), and TV.setChannel(). For an illustration, we show at the top of Fig. 1 that a service engineer sets out to discover interesting relationships hidden among services in the service repository with a general interest in smart home. The outcomes of mining service relationships in the smart home context might be quite surprising. For example, the TV service relates to a fridge service from three perspectives. The TV and fridge are located in the same home(i.e., spatial dependency). The fridge is generally accessed when the TV is on (i.e., temporal dependency). The TV and the fridge are sometimes accessed by the same person A (i.e, things relationships with people). From the discovered relationships between the TV service and the fridge service, the person A¡s behavior might be inferred, that he or she might like eating while watching TV. Another example is the relationship between a stove and an air-conditioning. The stove relates to the air-conditioning through environments. When the stove is in use, temperature is rising. Then the air-conditioning is invoked to adjust room temperature.

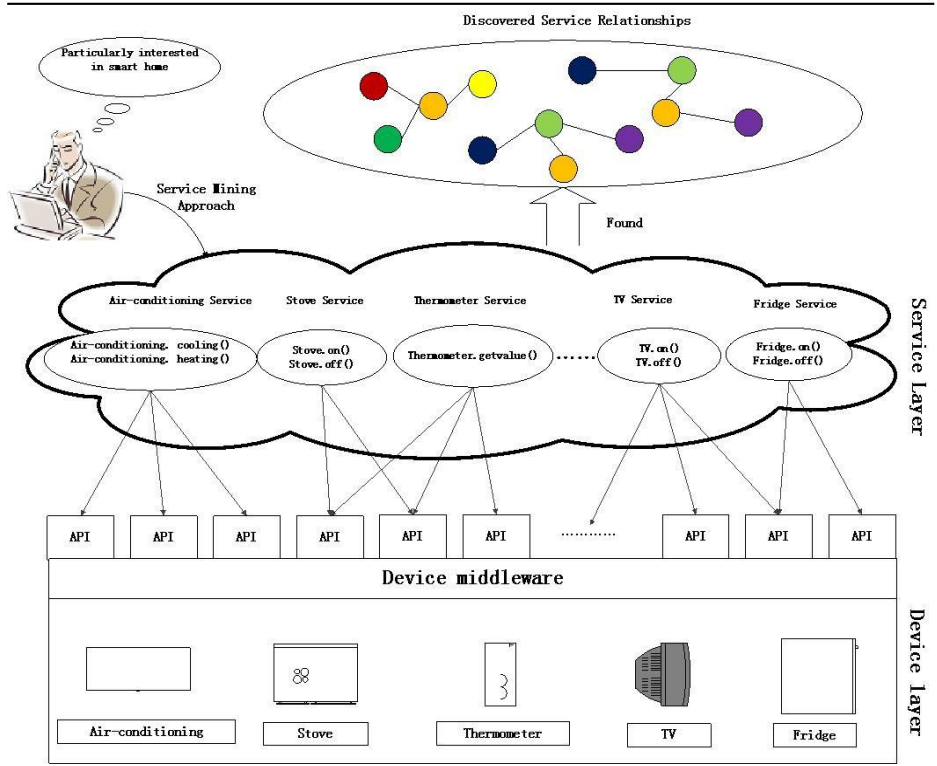

**Fig.1.** Service Mining Scenario

## 2   Related Work

**Service mining**

A Web service mining framework is proposed to discover service relationships [10]. The mechanism behind service relating to each other is that they are composable syntactically and semantically. This mechanism enables interesting and useful service composition to be discovered from bottom-up. In our work, the mechanism of service relationships is syntactic, semantic and context composability. In the service mining framework, interesting and usefulness evaluation mainly consists of two phases: objective evaluation and subjective evaluation [10]. The objective evaluation adopts a weighted function to sort out service compositions that exhibit high values of objective interestingness and usefulness. The basis of the objective interestingness and usefulness is operation similarity and domain correlation. The subjective evaluation is based on personal knowledge, belief, bias, and needs [10]. In our work, we adopt the objective evaluation concepts and subjective concepts proposed in the Web service mining framework [10]. Some research on service mining is based on service usage data or service description files (i.e., WSDL)[17][27][28][29][30]. An algorithm is proposed to discover service patterns by analyzing service usage data at user request, instance, and template levels [17]. Discovering similarity relationships among services (i.e., service clustering) is able to facilitate service discovery. For non-semantic Web services, the similarity calculation is based on WSDL files [27][28]. WSDL files together with tagging data are employed to cluster Web services [29][30]. These works make an implicit assumption that service relationships can be discovered indirectly through analyzing service usage data or service description files. One obvious limitation of these works is that service relationships cannot be discovered if there is no or little service usage data or description files available. In our work, the proposed service mining framework enables discovering service relationships to be carried out without service usage data.

**Service description**

One popular language for describing functional attributes of Web services is Web Service Description Language(WSDL)[18]. WSDL mainly provides syntactic descriptions of Web services with little support for semantic descriptions. To cater for semantic enabled services, WSDL is extended with semantic description capabilities [15]. A service ontology for describing Web services is proposed, which is specified using DAML+OIL language [15]. In our work, we extend the service ontology model proposed in [15] by incorporating context attributes descriptions, precondition and postcondition descriptions of operations.

**Context Modeling**

A hot research topic in ubiquitous computing is context-awareness [22][23][24][25] [26]. Services in ubiquitous computing need to be context-aware so that they can adapt themselves to dynamic situations. Ontology is one promising approach in modeling contexts. Ontology-based context model is proposed to describe context information of users [22][23].User¡s location, status, and temporal context are deduced by reasoning the ontology-based context model [22][23]. The ontology-based context model is proposed to address critical issues including formal context representation, knowledge sharing and logic based context reasoning [22][23][24]. The ontology-based context model has some limitations in reasoning user behaviors for lacking generality. One reason is that the ontology-based context model fails to capture the environment information, which is closely related to user behaviors. The other reason is that the ontology-based context model does not consider the relationships between users and physical things which can be used to deduce user behavior. In [25], an ontology is proposed to model home environments. The ontology model only considers the location of things. In [22], an ontology-based context model using OWL is proposed to define contexts in a semantic way. The ontology-based context model enables implicit contexts to be derived through context reasoning. The ontology-based context model is defined from a high level, resulting in low accuracy of context reasoning. It only considers people and location without considering environments. Those proposed ontology-based context models discussed above are context centered lacking generality. Thus these models have limitations in reasoning people behaviors. In our paper, the approach of capturing context information is service centered. We describe context information of services using ontology concepts.

## 3  Ontology-based Description of Services

Discovering service relationships requires the description of services so that services can understand and recognize each other. In the following, we propose a new service ontology model which is an extension to existing Web service ontology model (i.e., DAML+OIL[15]) incorporating context descriptions. We model the ontology using a directed graph (see Fig.2). Nodes represent the ontology concepts. Unfilled nodes refer to ontology concepts that have been clearly defined in [15]. Gray nodes refer to extended ontology concepts. Edges denote relationships between the ontology concepts. For example, the edge operation → precondition states that an operation has no or multiple preconditions.

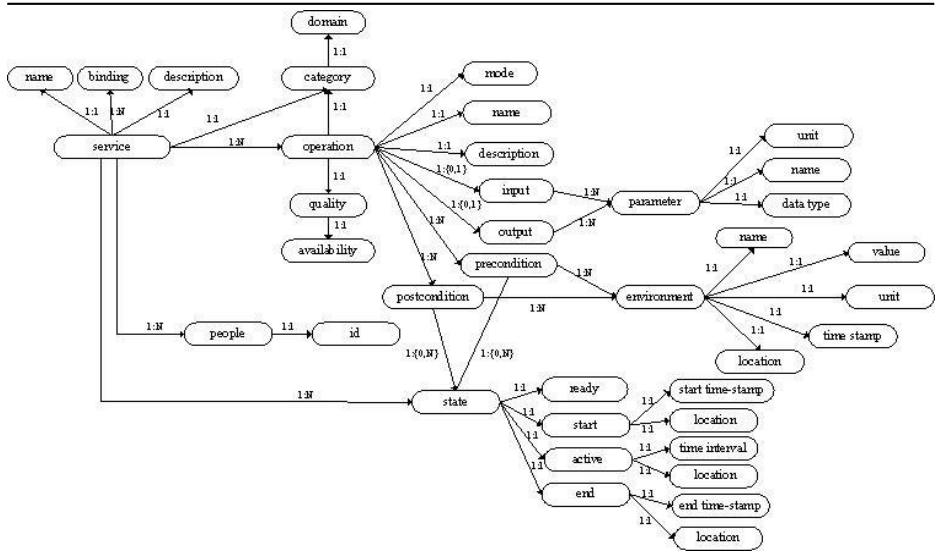

**Fig.2.** Ontology-based Description of Services

## 3.1   Environment

Environments refers to physical environments such as light, voice, and temperature etc. which can be measured by sensors. Environments are important factors in analyzing service relationships. On the one hand, the effects of service execution may impact environments. For example, the effects of the stove service execution make the room temperature increase. On the other hand, environments may impact services. For example, when the room temperature is higher than a preferred degree, the air-conditioning services will be invoked to cool the room. Therefore, the air-conditioner service relates to the stove service via room temperature.We identify five concepts to describe environments, namely, name, value, unit, time-stamp, and location. We formally define environment as follows:

*Definition 1:* Environment. The environment $Env_i$ is defined as a tuple ⟨ *name*, (*val*,*uni*,*ts_i*,*loc_i*) ⟩, where:

• *name* is the unique identifier of $Env_i$.

•(*val*,*uni*,*ts_i*,*loc_i*) shows the recorded data value *val*, unit *uni*, time-stamp *ts_i* and location *loc_i* that are related to environment $Env_i$ . In our paper, location is uniformly defined as a GPS point with a user-defined spatial radius *r*.

For simplicity, we use $v_t$ to refer to the environment value at time point *t* . For example, the room temperature is 25 degree at 3pm, which can be formalized as ⟨temperature, 25, degree, 3pm, room⟩.

### 3.2  State

We define state as externally observable property of services. We identify four types of service states, namely, *ready*, *start*, *active*, and *end*. The state can be described based on spatio-temporal features. We formally define the *state* as follows:

*Definition 2:* State. The state $Sta_i$ is defined as a tuple ⟨*ready*, *start*, *active*, *end*⟩ where:

- *ready*: the service is in the ready state if the request for invoking the service has not been made.
- *start*: the start state means that the service execution has been initiated. The start state is defined by a tuple ⟨$st_i$, $loc_i$⟩ where $st_i$, and $loc_i$ are the start time stamp and location of initiating the service, respectively.
- *active*: the service is in the *active* state if the service is being executing. The active state is defined by a tuple ⟨$tin_i$, $loc_i$⟩ where $tin_i$ and $loc_i$ are the time interval and location of the service execution, respectively; and $tin_i = et_i st_i$.
- *end*: the service is in the end state if the service execution is terminated. The end state is defined by a tuple ⟨$et_i$, $loc_i$⟩ where $et_i$ and $loc_i$ are the end time stamp and location of terminating the service, respectively.

The *start* and *end* states of services are instantaneous, while the *active* and *ready* states sustain a period of time. For example, the TV is on from 9 am to 12 am in the bedroom. In this example, the TV is in *active* state for 3 hours in the bedroom.

### 3.3  Precondition and Postcondition

Service operation execution may require going through predefined environments and states (i.e., precondition). The effects of service operation execution may impact environments and alter service states (i.e, postcondition). We define the precondition of an operation as a set of conditions that must be satisfied before executing the operation. The postcondition of an operation is defined as a set of effects resulting from executing an operation. We formalize precondition and postcondition as follows:

*Definition 2:* Precondition and Postcondition. The precondition $Pre_i$ is defined as a tuple ⟨($Sta_1$, $Sta_2$¡ $Sta_i$), ($Env_1$, $Env_2$¡ $Env_i$)⟩ where:

- *Sta$_i$* is the required *state* before executing an operation (cf. Definition 2).
- *Env$_i$* is the required *environment* before executing an operation (cf. Definition 1).

The postcondition *Pos$_i$* is defined as a tuple h (*Sta$_1$*, *Sta$_2$*¡ *Sta$_i$*), (*Env$_1$*, *Env$_2$*¡ *Env$_i$*)i where:

- *Sta$_i$* is the effect state after executing an operation (cf. Definition 2).
- *Env$_i$* is the effect environment after executing an operation (cf. Definition 1).

For example, the operation air-conditioningr.setTemperature() is executed before the precondition h air-conditioning.state=active, temperature¿28 degree i is satisfied. The postcondition of executing this is that the htemperature =25 degreei.

### 3.4   Operation

Each operation is identified by a category, a mode, a name, a description, input and/or output messages, and quality concepts [15]. These concepts definitions refer to [15].We extend operation definition by incorporating precondition and a postcondition. We define an operation as follows:

Definition 4: Operation. An operation *Ope$_i$* is defined by a tuple h*name*, *des*, *Cat$_i$*, *Mod$_i$*, *Inp$_i$*, *Out$_i$*, *Qua$_i$*, *Pre$_i$*, *Pos$_i$*i where:

- *name*, *des*, *Cat$_i$*, *Mod$_i$*, *Inp$_i$*, *Out$_i$*, *Qua$_i$* refer to a name, a text description, categories, modes, input messages, output messages, and qualities concepts, respectively[15]
- *Pre$_i$* gives the preconditions of an operation (cf. Definition 3).
- *Pos$_i$* gives the postconditions of an operation (cf.Definition 3).

For example, the operation air-conditioning.setTemperature() is described as the tupleh ¡setTemperature¡, ¡This operation can adjust temperature to a preferred degree¡, ¡appliance¡, ¡one-way¡, (temperature, int, degree), (none), ¡availability=true¡, (air-conditioner.state=active, temperaturei28 degree), (temperature =25 degree) i.

### 3.5 People

People can be either an individual or groups. We define people as follows:
Definition 5: People. *Peo$_i$* is defined as a tuple ⟨id⟩ where

- *id* is the unique identifier of people.

### 3.6 Service

A service is identified by a name, a text description, bindings, categories which are defined referring to [15]. We extend the service definition by incorporating state and people and modifying the operation. Definition 6: Service. A service $S_i$ is defined by a tuple ⟨*name*, *des*, *bind$_i$*, *Cat$_i$*, *Ope$_i$*, *Sta$_i$*, *Peo$_i$*⟩ where:

- *name*, and *des* are a name and a text summary about the service features, respectively.
- *bind$_i$* is a set of binding protocols supported by $S_i$.
- *Cat$_i$* is a set of categories that $S_i$ belongs to.
- *Ope$_i$* is a set of operations provided by $S_i$ (cf. Definition 4).
- *Sta$_i$* is a set of states that $S_i$ represents (cf. Definition 2).
- *Peo$_i$* is people who consume the service $S_i$ (cf. Definition 5).

For example, the air-conditioner service can be described by a tuple ⟨¡airconditioner service¡, ¡This service is used for adjusting room temperature¡, (SOAP), (appliance), (air-conditioner.On(), air-conditioner.Off(), air-conditioner.setTemperature(), air-conditioner.Ventilation(), air-conditioner.setTime()), (ready, start, active, end), (Nancy)⟩.

## 4 Evaluation

Not all service composition leads discovered in the service recognition phase are necessarily interesting. Evaluation measures are needed to filter out uninteresting service composition leads. The evaluation measures consists of two steps: objective evaluation and subjective evaluation.

### 4.1 Objective Evaluation

For performance reasons, the objective evaluation consists of two steps. The first step is *Correlation Degree*(*CD*) filtering. The *CD* gives the relationship strength between two services. We define *Correlation Degree*(*CD*) in Eq.(4).

*CD*(*S_i*,*S_k*) = $\eta_1$·*State*(*S_i*,*S_k*)+$\eta_2$·*Env*(*S_i*,*S_k*)+$\eta_3$·*Peo*(*S_i*,*S_k*)+$\eta_4$·*Ope*(*S_i*,*S_k*)

(4)

Where $\eta_1$ +$\eta_2$ +$\eta_3$ +$\eta_4$ = 1,*State*(*S_i*,*S_k*),*Env*(*S_i*,*S_k*),*Peo*(*S_i*,*S_k*),*Ope*(*S_i*,*S_k*) are binary values returned from returned from service recognition algorithms. We define a *correlation threshold*($\zeta$) which gives the minimum value allowed for the *CD*. If *CD*(*S_i*,*S_k*) ≥ $\zeta$, then leads of composed services *S_i* and *S_k* are selected for further evaluation. Otherwise and are filtered out.

The second step is *interestingness* evaluation. Since *CD* is a ¡coarse-grained¡ evaluation. The number of service composition leads sorted out from the *CD* evaluation phase may still be large. Therefore, service composition leads need a further evaluation. We first give the concepts of ,*Actionability*(*Act*) , *Domain Correlation*(*DC*), and *Diversity*(*Div*), and then give the *interestingness* definition

*Actionability*(*Act*). *Act* defined as a binary (i.e., 1 for actionable, 0 for nonactionable) representing whether the discovered service composition leads can be verified through simulation [10]. A service composition lead that cannot be verified is considered uninteresting.

*Novelty*. *Nov* is defined as a binary (i.e., 1 for novel, 0 for known). The source of novelty information may be a registry that keeps tacks of known service compositions [10]. A known service composition lead is considered uninteresting.

*Domain Correlation*. *DC* measures the relevance of two domains that services *S_i* and *S_k* belong to, respectively. As defined in Eq.2, *Ont*(*S_i*) denotes the domain ontology that service *S_i* refers to. Therefore, the domain correlation of *S_i* and *S_k* is equivalent to the ontology correlation of *Ont*(*S_i*) and *Ont*(*S_k*). We formally define *DC* in Eq.(5).

$$DC(S_i, S_k) = e^{-\frac{1}{\lambda_0 \cdot \{sim(Ont(S_i), Ont(S_k))+1\}}}$$

(5)

Where

*Sim*(*Ont*(*S_i*),*Ont*(*S_k*)) =

$$w_1 \cdot \left(\frac{|N_{pre}(S_i)| \cap |N_{pre}(S_k)|}{|N_{pre}(S_i)| \cup |N_{pre}(S_k)|}\right) + w_2 \cdot \left(\frac{|N_{pos}(S_i)| \cap |N_{pos}(S_k)|}{|N_{pos}(S_i)| \cup |N_{pos}(S_k)|}\right)$$
$$+ w_3 \cdot \left(\frac{|N_{in}(S_i)| \cap |N_{in}(S_k)|}{|N_{in}(S_i)| \cup |N_{in}(S_k)|}\right) + w_4 \cdot \left(\frac{|N_{out}(S_i)| \cap |N_{out}(S_k)|}{|N_{out}(S_i)| \cup |N_{out}(S_k)|}\right)$$

Where *S_i*,*S_k* ∈ *ML*,*w_i*(*i* = 1,2,3,4) is weight such that *w_i* ∈ [0,1] and $\sum_1^4 w_i = 1$. $N_{pre}(S_i), N_{pos}(S_i), N_{in}(S_i)$ and *N_out*(*S_i*) refer to the set of preconditions, postconditions, input parameters and output parameters concepts, respectively. Operators ∩ and ∪ are based on ontological overlap and union of two concepts. When *Sim*(*Ont*(*S_i*),*Ont*(*S_k*)) = 0, the correlation between two domains is assigned an initial value $r_0 = e^{-\frac{1}{\lambda_0}}$. E.q 5 shows that $DC(S_i,S_k)$ approaches 1 as $Sim(Ont(S_i),Ont(S_k))$ increases. We define *Div* as the multiplicative inverse of the domain correlation. We bound the maximum value of *Div* to 1. The *Div* is formally defined as follows.

$$Div = \frac{r_0}{DC(S_i, S_k)} \quad (6)$$

Consequently, the objective interestingness function is defined as follows. *Interestingness* = *Act*·$w_1$+*Nov*·$w_2$+*Div*·$w_3$ where $S_i, S_k \in ML, w_1, w_2, and w_3$ are weights for Actionability, Novelty, and Diversity respectively, $w_i \in [0,1] (i = 1,2,3)$, and $\sum_1^3 w_i = 1$. We define an interesting threshold($\xi$) which gives the minimum value allowed for the . If the value of *interestingness* $\geq \xi$, then leads of composed services are considered interesting. Otherwise the service composition leads are considered uninteresting.

### 4.2 Subjective Evaluation

The objective evaluation narrows down large numbers of discovered service composition leads to a smaller number of candidates through *Correlation Degree* and *interestingness* function. In addition to the objective evaluation, users can evaluate the interestingness of service composition leads subjectively. Subjective evaluation is based on personal knowledge, behaviors, habits, and needs. Thus, the interestingness of the same service composition lead may be distinct for different people. For example, a home doctor may think the TV-fridge composition lead is interesting because it reflects the eating habit of patients who suffer from a digest problem. While the TV-fridge composition lead may be uninteresting for healthy people. Based on the subjective evaluation, the service engineer makes the ultimate decision as to whether the discovered service composition leads is interesting.

## 5 Experiment Results

We study the effects of variables listed in Table 1 on the total number of discovered service composition leads, the average values of Correlation Degree, the number of interesting service composition leads. the average values of interestingness. We run on our experiment on a 1.6 GHZ AMD processor and 4 GB RAM under Windows 7. To the best of our knowledge, there is no services test case for IoT to be used for experimental purposes. Therefore, we evaluate the variables effects using simulated services. For each service, we randomly generate its input/output parameters such that the number of these parameters uniformly falls in the range of 0 to 5. Each of these parameters is assigned to an agent. For simplicity, we only consider the exact input/output parameter data type match. To simulate the precondition and postconditions, the precondition and postcondition is symbolically given 0 to 3 parameters which uniformly falls in the range of 0 to 9. To simulate temporal attributes, we assign a time interval for each service¡s temporal attributes. We use the overlap of two time intervals to simulate the temporal dependency of two services. For simplicity, we assume the location regarding each service is the same. Thus the spatial dependency of two services can be ignored. Since both actionability and novelty are boolean variables, the two variables are assumed to be true. For the people attributes of services, we assign an alphabet randomly to each service¡s people attributes. Finally, we use an interestingness threshold of 0.6 to determine whether a service composition lead is interesting. The interesting threshold can be changed according to acceptable interestingness expectations.

| Variable | Value or Rang |
|---|---|
| Number of operations per service | 1-5 |
| Input parameters per service | 0-5 |
| Output parameters per service | 0-5 |
| Number of pre/pos-condition per operation | 0-3 |
| Range of pre/pos-condition per operation | 0-9 |
| Temporal range | 0-24 |
| Actionability/novelty | 1/1 |
| $\eta_1/\eta_2/\eta_3/\eta_4$ | 0.1/0.2/0.3/0.4 |
| $w_1/w_2/w_3$ | 0.3/0.3/0.4 |
| $r_0$ | 0.1 |
| $\zeta$ | 0.5 |
| $\xi$ | 0.7 |

**Table 1.** Experiment Settings

Service mining for Internet of Things    154. Viani, Federico, et al. "Wireless architectures for heterogeneous sensing in smarthome applications: Concepts and real implementation." Proceedings of the IEEE 101.11 (2013): 2381-2396.
5. Philipose, Matthai, et al. "Inferring activities from interactions with objects." Pervasive Computing, IEEE 3.4 (2004): 50-57.
6. Han, Jiawei, Micheline Kamber, and Jian Pei. Data mining: concepts and techniques: concepts and techniques. Elsevier, 2011
7. Strunk, Anja. "QoS-aware service composition: A survey." Web Services (ECOWS),2010 IEEE 8th European Conference on. IEEE, 2010.
8. Dey, Anind K., Gregory D. Abowd, and Daniel Salber. "A context-based infrastructure for smart environments." Managing Interactions in Smart Environments. Springer London, 2000. 114-128.
9. Neiat, Azadeh Ghari, et al. "Failure-proof spatio-temporal composition of sensorcloud services." Service-Oriented Computing. Springer Berlin Heidelberg, 2014. 368377.
10. Zheng, George, and Athman Bouguettaya. "Service mining on the web." ServicesComputing, IEEE Transactions on 2.1 (2009): 65-78.
11. Yao, Lina, et al. "Things of Interest Recommendation by Leveraging HeterogeneousRelations in the Internet of Things."
12. L. Tan and N. Wang, ¡Future internet: The internet of things,¡ in Proc.3rd Int.Conf. Adv. Comput. Theory Eng. (ICACTE), Chengdu, China,Aug. 20C22, 2010, pp. V5-376CV5-380.
13. Yao, Lina, et al. "A model for discovering correlations of ubiquitous things." DataMining (ICDM), 2013 IEEE 13th International Conference on. IEEE, 2013.
14. Atzori, Luigi, Antonio Iera, and Giacomo Morabito. "The internet of things: Asurvey." Computer networks 54.15 (2010): 2787-2805.
15. Medjahed, Brahim, Athman Bouguettaya, and Ahmed K. Elmagarmid. "Composing web services on the semantic web."The VLDB Journal¡aThe International Journal on Very Large Data Bases 12.4 (2003): 333-351.
16. Neiat, Azadeh Ghari, Athman Bouguettaya, and Timos Sellis. "Spatio-TemporalComposition of Crowdsourced Services."Service-Oriented Computing. Springer Berlin Heidelberg, 2015. 373-382.
17. Liang, Qianhui Althea, et al. "Service pattern discovery of web service miningin web service registry-repository."e-Business Engineering, 2006. ICEBE'06. IEEE International Conference on. IEEE, 2006.
18. Christensen, Erik, et al. "Web services description language (WSDL) 1.1." (2001).
19. Burstein, Mark, et al. "OWL-S: Semantic markup for web services." W3C MemberSubmission(2004).
20. Roman, Dumitru, et al. "Web service modeling ontology."Applied ontology1.1 (2005): 77-106.
21. Yao, Lina, et al. "A model for discovering correlations of ubiquitous things." DataMining (ICDM), 2013 IEEE 13th International Conference on. IEEE, 2013
22. Wang, Xiao Hang, et al. "Ontology based context modeling and reasoning usingOWL." Pervasive Computing and Communications Workshops, 2004. Proceedings of the Second IEEE Annual Conference on. Ieee, 2004.
23. Zhang, Daqing, Tao Gu, and Xiaohang Wang. "Enabling context-aware smarthome with semantic web technologies." International Journal of Human-friendly Welfare Robotic Systems 6.4 (2005): 12-20.